\documentclass[runningheads]{llncs}

\PassOptionsToPackage{numbers, compress}{natbib}

\usepackage{siunitx}

\usepackage{algorithm}
\usepackage[noend]{algpseudocode}
\usepackage{graphicx,xfp}
\usepackage{multicol}
\usepackage{multirow}
\usepackage{color}
\usepackage{xcolor}
\usepackage{booktabs}
\usepackage{nicefrac}
\usepackage{amsmath}
\usepackage{array}
\newcolumntype{P}[1]{>{\centering\arraybackslash}p{#1}}

\usepackage{pifont}%

\usepackage{textcomp}
\usepackage{sidecap}
\usepackage{graphicx,xfp}
\usepackage{amssymb}

\usepackage{amsthm}
\theoremstyle{definition}
\theoremstyle{remark}

\theoremstyle{plain}
\theoremstyle{definition}

\usepackage[colorlinks]{hyperref}

\usepackage{xfp}

\usepackage{tikz}
\usepackage{tikzscale}
\usepackage{xspace}
\usepackage{pgfplots}
\usepgfplotslibrary{groupplots}
\pgfplotsset{compat=1.16}
\usepackage{xparse}
\NewDocumentCommand{\Log}{o}{%
  \IfNoValueTF{#1}{}{{}^{#1}\!}\log}%

\usepackage{stmaryrd}

\usepackage{subcaption}

\begin{document}




\title{Sequence-Based Nanobody-Antigen Binding Prediction}
\titlerunning{Sequence-Based Nanobody-Antigen Binding Prediction}

\author{Usama Sardar\inst{1}$^*$ \and
Sarwan Ali\inst{2}$^*$ \and
Muhammad Sohaib Ayub\inst{1} \and
Muhammad Shoaib\inst{1} \and
Khurram Bashir\inst{1} \and
Imdad Ullah Khan\inst{1} \and
Murray Patterson\inst{2}
}
\authorrunning{U. Sardar et al.}
%
\institute{Lahore University of Management Sciences, Lahore, Pakistan \\ 
\email{usamasardar2022@gmail.com, \{15030039,mshoaib,khurram.bashir,imdad.khan\}@lums.edu.pk}
\\
\and
Georgia State University, Atlanta GA, USA \\
\email{\{sali85,mpatterson30\}@gsu.edu}
\\
$^*$ Equal Contribution
}

\raggedbottom
\maketitle

\begin{abstract}
Nanobodies (Nb) are monomeric heavy-chain fragments derived from heavy-chain only antibodies naturally found in Camelids and Sharks. Their considerably small size ($\sim$3-4 nm; 13 kDa) and favorable biophysical properties make them attractive targets for recombinant production. Furthermore, their unique ability to bind selectively to specific antigens, such as toxins, chemicals, bacteria, and viruses, makes them powerful tools in cell biology, structural biology, medical diagnostics, and future therapeutic agents in treating cancer and other serious illnesses. However, a critical challenge in nanobodies production is the unavailability of nanobodies for a majority of antigens. Although some computational methods have been proposed to screen potential nanobodies for given target antigens, their practical application is highly restricted due to their reliance on 3D structures. Moreover, predicting nanobody-antigen interactions (binding) is a time-consuming and labor-intensive task. This study aims to develop a machine-learning method to predict Nanobody-Antigen binding solely based on the sequence data. We curated a comprehensive dataset of Nanobody-Antigen binding and non-binding data and devised an embedding method based on gapped $k$-mers to predict binding based only on sequences of nanobody and antigen. Our approach achieves up to $90\%$ accuracy in binding prediction and is significantly more efficient compared to the widely-used computational docking technique.




\end{abstract}

\section{Introduction}

Nanobodies (Nbs) are single-domain antibodies (sdAb), derived from heavy-chain only antibodies naturally occurring in Camelids and Sharks. They represent a unique class of proteins/antibodies having a molecular weight of 12–15 kDa, that combine the advantageous characteristics of conventional antibodies with desirable attributes of small-molecule drugs. Nbs are remarkably adaptable to various applications and offer several advantages over conventional antibodies~\cite{cortez2004efficient}. Every Nb contains the distinct structural and functional properties found in naturally-occurring heavy-chain antibodies. They have a naturally low potential for causing immune responses and exhibit high similarity to variable region of heavy chain (VH) in human antibodies, making them excellently suited for therapeutic and diagnostic applications. Due to their small size, unique structure and high stability, Nbs can access targets that are beyond the reach of conventional antibodies and small-molecule drugs~\cite{revets2005nanobodies,deffar2009nanobodies}. Nbs Structure prediction and modeling are still challenging tasks~\cite {cohen2022nanonet,valdes2023structural}. Hundreds of Nb crystallographic structures have been deposited in the Protein Data Bank (PDB)~\cite{berman2000protein,burley2019rcsb}. Despite this, the current representation falls short of capturing the vast structural and sequence diversity observed in Nb hypervariable loops. Moreover, Nbs display a greater range of conformational variations, lengths, and sequence variability in their CDR3 compared to antibodies~\cite{mitchell2018analysis}. This makes modeling and prediction of their 3D structure more complex.

Machine Learning (ML) plays a crucial role in predicting nanobody-antigen(Nb-Ag) interactions. ML offers a powerful and effective approach to analyzing and comprehending complex patterns within extensive datasets~\cite{farhan2017efficient}. Traditional non-computational methods for determining Nb-Ag interactions can be both costly and time-consuming. ML provides a faster and more cost-effective alternative, enabling scientists to prioritize potential nanobodies candidates for further research~\cite{tam2021nbx}. The large amount of training data that is easily accessible when utilizing sequence-based ML techniques is advantageous. Even though there is an increasing amount of data on protein structures, most Nb-Ag sequences still lack validated structural details, even though the number of protein sequence entries is still rising quickly~\cite{schwede2013protein,hou2021serendip}.

 ML algorithms are capable of handling a vast amount of data, encompassing nanobodies and antigen sequences, structural information, and experimental binding data. Through the examination of this data, ML algorithms can find complex relationships and patterns that might be hard for people to see on their own. These methods can automatically extract relevant details from raw data, such as structural information or amino acid sequences. This feature extraction method makes it easier to spot important molecular traits that influence antibody and Nb-Ag interaction~\cite {myung2022csm,yang2023area}. ML models can gain knowledge from large databases of training samples and produce precise predictions. By being trained on known Nb-Ag binding data, these models may understand the underlying principles and patterns regulating binding interactions. This allows them to make accurate predictions for previously undiscovered  Nb-Ag combinations. Traditional experimental techniques, on the other hand, need a lot of time and money to determine  Nb-Ag binding. To arrange possible antibody candidates for further investigation, ML offers a quicker and more economical alternative~\cite{ramon2023abnativ,ye2022prediction,yang2023area,miller2023learned}.

Nanobody-antigen binding play a significant role in the immune response. By predicting and studying these bindings, researchers can gain insights into how nanobodies recognize and neutralize specific antigens. This understanding is fundamental for elucidating immune mechanisms and developing strategies for diagnostics to combat infectious diseases, autoimmune diseases, and cancer. Predicting Nb-ag binding can aid in discovering and engineering nanobodies with desired properties. Binding helps identify the most essential antigens for vaccine development, effective vaccine formulations, and understanding the mechanisms of immune protection. 

Predicting binding interactions enables the selection of highly specific nanobodies and antigens for accurate and sensitive diagnostic tests.  Binding can guide the rational design of therapeutic nanobodies or nanobodies-based drugs and diagnostic tests. Scientists can modify or engineer nanobodies to improve their affinity, selectivity, and therapeutic potential by understanding the binding interactions between nanobodies and their target antigens. This approach can be applied in areas such as diagnostic tests and cancer immunotherapy, where nanobodies are designed to target specific tumor antigens~\cite {sormanni2015rational}.

 The input in our study is nanobody sequences and antigen sequences  and the output is binding/docking score (Yes/No).In this paper, we trained ML on nanobodies and antigen sequences extracted from the single-domain antibody (sdAb) database to determine binding. We make the following contributions.




\begin{itemize}
    \item We have performed the comparison of various ML approaches for predicting nanobody-antigen binding from sequences only. 
    \item We evaluated the impact of various sequence features (e.g., isoelectric point, hydrophilicity) on the prediction accuracy of ML models.
    \item We have curated a dataset of nanobody-antigen pairs for training and testing machine-learning models and made it publicly available for further research.
\end{itemize}

The rest of the paper is organized as follows: Section~\ref{related_work} explores the existing research and highlights the research gaps. Section~\ref{proposed_approach} explains the data collection, feature extraction, and data visualization. The proposed embedding is discussed in Section~\ref{sec_embedding_generation}. Section~\ref{experimental_setup} describes the ML models and evaluation metrics. We discuss our results in Section~\ref{results_discussion}. Finally, we provide the conclusion and future directions in Section~\ref{sec_conclusion}.

\section{Related Work} \label{related_work}


Antibodies (Abs) are crucial tools in biological research and the biopharmaceutical industry due to their exceptional binding specificity and strong affinity for target antigens. The effectiveness of the immune system directly reflects the diversity of antigens against which specific tightly binding `B-lymphocyte antigen receptors (BCRs) can be generated. The vast range of binding specificity is achieved through sequence variations in the heavy chain (VH) and light chain (VL), resulting in an estimated diversity of BCRs in humans~\cite{mitchell2018comparative} that surpasses the population size of B-lymphocytes in an individual. However, it is still unclear how this immense sequence diversity translates into antigen specificity. Although not every unique combination of VH-VL sequences leads to a distinct binding specificity, predicting the number and positions of amino acid mutations required to change binding specificity has proven challenging~\cite{peng2014origins}. 

\begin{figure}[h!]
    \centering
    \includegraphics[width=0.5\textwidth]{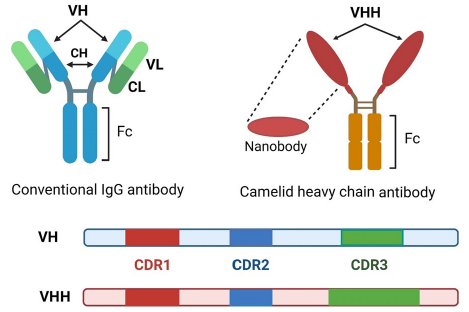}
    \caption{Nanobody vs Antibody (Figure taken from~\cite{hosseindokht2021nanobodies})}
    \label{fig:Ab_vs_Nb}
\end{figure}

A more manageable system is offered by heavy-chain antibodies found in camelid species like camels, llamas, and alpacas, where the light chain is absent, as shown in Figure~\ref{fig:Ab_vs_Nb}. These antibodies, known as nanobodies (Nbs), consist of  an isolated variable VHH domain that is about ten times smaller than conventional Abs but retains comparable binding specificity~\cite{muyldermans2013nanobodies}.

Both Nbs and Abs face the fundamental challenge of deciphering the molecular code that links amino acid sequence, particularly the choice of paratope residues, to the binding specificity of the folded molecule. In regular Abs, the paratope is situated at the interface of the VH and VL domains, typically comprising residues from up to six distinct hypervariable loop regions. The VH and VL domains can dock together in various ways, allowing the antibody to maximize the diversity of potential antigen-binding surfaces. In contrast, the Nb paratope is entirely contained within the VHH domain, significantly limiting the range of possible antigen-binding surfaces without seemingly affecting the diversity of resulting binding specificities. Indeed, Nbs typically bind their target antigens with affinities comparable to classical monoclonal Abs~\cite{mitchell2018comparative}.


Several studies have been conducted to generate antibodies/nanobodies using non-computational methods. Experimental techniques such as hybridoma technology and phage display~\cite{rossant2014phage} are used to generate specific antibodies/nanobodies, but these have limitations and challenges. Hybridoma technology involves immunizing animals with an antigen and fusing B cells from the immunized animal with cancer cells to create hybridoma cells that produce specific antibodies. However, this method raises ethical concerns due to animal cruelty and is time-consuming, labor-intensive, and limited in antibody/nanobody diversity. On the other hand, phage display utilizes bacteriophages to display antibody fragments, but it also has time-consuming rounds of selection and amplification, labor-intensive requirements, and high costs.


Several ML methods are used to predict nanobody-antigen binding. Sequence-based methods utilize amino acid sequences, extracting features such as physio-chemical properties, sequence motifs, and sequence profiles~\cite{tam2021nbx}. These features are then used as input for machine learning algorithms like support vector machines (SVM), random forests, or neural networks. Structure-based methods employ three-dimensional structures obtained from experimental techniques like X-ray crystallography or homology modeling. Structural features like solvent accessibility, electrostatic potential, or shape complementarity are extracted and fed into machine-learning models. Hybrid methods combine sequence-based and structure-based features, integrating both sequence and structural information to capture a broader range of characteristics. Deep learning methods, such as convolutional neural networks (CNN) and recurrent neural networks (RNN), learn complex patterns and relationships from large datasets, including sequence and structural information, for accurate predictions~\cite{cohen2022nanonet}. Docking-based methods use molecular docking algorithms to predict the binding orientation and affinity by calculating a binding score based on the optimal spatial arrangement of the interacting molecules~\cite{ramon2023abnativ}.

\section{Proposed Approach} \label{proposed_approach}
The proposed pipeline comprised different steps including data collection, Nb-Ag sequence analysis, numerical embedding generation, and optimal feature extraction from the Nb and Ag sequences. We discuss each step in detail.

\subsection{Data Collection}
We collected $47$ Ag sequences from UniProt~\footnote{\url{https://www.uniprot.org/}}, and for each we collected all binding Nbs from Single Domain Antibody Database\footnote{\url{http://www.sdab-db.ca/}}, which are total $365$, as shown in Table~\ref{tbl_seq_lenghts} along with the basic statistics for the length of the antigen sequences including average, minimum, and maximum lengths, etc. Basica summary of the number of nanobodies binding to antigens is given in Table~\ref{tbl_freq_seq}


 \begin{table}[h!]
     \centering
     \begin{tabular}{p{2cm}p{1.5cm}p{1.5cm}p{1.5cm}p{1.5cm}p{1.5cm}p{1.5cm}}
     \toprule
     & & \multicolumn{5}{c}{Sequence Length Statistics} \\
     \cmidrule{3-7}
          Type & Count & Mean & Min &  Max &  Std. Dev. & Median \\
          \midrule
           Antigens & $47$ & $671.51$ & $158$ & $1816$ & $421.24$ & $480$ \\
           Nanobodies & $365$ & $122.84$ & $104$ & $175$ & $8.87$ & $123$ \\
          \bottomrule
     \end{tabular}
     \caption{Sequence length statistics for antigen and nanobody sequences.} 
     \label{tbl_seq_lenghts}
 \end{table}

 \begin{table}[h!]
     \centering
     \begin{tabular}{p{1cm}lp{1.4cm}p{1.4cm}p{1.4cm}p{1.4cm}p{1.4cm}}
     \toprule
          Type & & Mean & Min &  Max &  Std. Dev. & Median \\
          \midrule
          \multicolumn{2}{l}{Nanobodies in each antigen} & $7.77$ & $1$ & $36$ & $9.28$ & $4$ \\
          \bottomrule
     \end{tabular}
     \caption{Statistics for nanobody sequences binding to each antigen.} 
     \label{tbl_freq_seq}
 \end{table}

\subsection{Features Extracted From Sequences}\label{sec_seq_features}

We performed basic protein sequence analysis using the `bioPython' package\footnote{\url{https://biopython.org/docs/dev/api/Bio.SeqUtils.ProtParam.html}} on each nanobody and antigen sequence to determine their features. 
These features include charge at pH,	Grand Average of Hydropathy (GRAVY), molecular weight, aromaticity, instability index, isoelectric point, secondary structure fraction (helix, turn, and sheet), and molar extinction coefficient (reduced and oxidized).

\paragraph{\textbf{Charge at pH: }} The charge of a protein at a pH is determined by presence of charged amino acids (aspartic acid, glutamic acid, lysine, arginine, histidine, and cysteine) and their ionization state. These amino acids gain or lose protons at different pH values, resulting in a net charge. The charge at pH affects the protein's solubility, interaction with other molecules, and biological function.

\paragraph{\textbf{Grand Average of Hydropathy (GRAVY): }} GRAVY~\cite{kyte1982simple} measures the overall hydrophobicity or hydrophilicity of a protein, calculated by averaging the hydropathy values of its amino acids. Positive GRAVY values indicate hydrophobic, while negative values represent hydrophilic regions and provide insights into protein stability, membrane interactions, and protein-protein interactions.

\paragraph{\textbf{Molecular weight: }} Molecular weight refers to the sum of the atomic weights of all atoms in a protein molecule. It is calculated based on the amino acid composition of the protein sequence. Molecular weight impacts various protein properties, such as protein folding, thermal stability, and mobility, and is crucial for protein identification, characterization, and quantification.


\paragraph{\textbf{Instability index: }} The instability index~\cite{guruprasad1990correlation} is a measure of the propensity of a protein to undergo degradation or unfold. Higher instability index values indicate increased susceptibility to degradation and decreased protein stability. The index is useful for evaluating protein expression, protein engineering, and predicting potential regions of protein instability.

\paragraph{\textbf{Isoelectric point: }} The isoelectric point (pI) is the pH at which a protein has a net charge of zero. It is determined by the presence of charged amino acids and their ionization states. The pI influences protein solubility, crystallization, and electrophoretic mobility. Knowledge of the pI is crucial for protein purification, protein characterization, and protein separation techniques based on charge.

\paragraph{\textbf{Secondary structure fraction (helix, turn, and sheet): }} The secondary structure fraction~\cite{haimov2016closer,hutchinson1994revised,kim1993thermodynamic} refers to the proportions or percentages of different secondary structure elements (helices, turns, and sheets) in a protein sequence. These elements are determined by the pattern of hydrogen bonds between amino acids. The secondary structure fraction provides insights into the protein's folding, stability, and functional properties. Different secondary structure fractions contribute to the unique 3D structure and biological function of the protein.

\paragraph{\textbf{Molar extinction coefficient (reduced and oxidized): }} The molar extinction coefficient is a measure of the ability of a molecule to absorb light at a specific wavelength. It quantifies the efficiency of light absorption by the molecule. The molar extinction coefficients can be different for reduced and oxidized forms of a protein due to changes in the chromophores. These coefficients are useful for protein quantification, monitoring protein folding/unfolding, and studying protein-protein interactions.

\begin{remark}
  We compute classification results with and without these features for all embedding methods, including the proposed and baseline methods.  
\end{remark}

\subsection{Obtaining Non-Binding Nb-Ag Pairs}

We created the proximity matrix of these sequences using Clustal Omega~\footnote{\url{https://www.ebi.ac.uk/Tools/msa/}} to evaluate the pairwise edit distance between antigens and nanobodies. This pairwise distance is used to identify further binding pairs and non-binding pairs of antigens and nanobodies. 
There are three pairs of antigens with very high similarity (distance $< .25$), namely, 
antigen green fluorescent protein (GFP) and Superfolder green fluorescent protein (sfGFP), (pairwise distance $= 0.05042$), RAC-gamma serine/threonine-protein kinase pleckstrin homology domain (Akt3PH) and RAC-alpha serine/threonine-protein kinase pleckstrin homology domain (Akt1PH) (pairwise distance $= 0.19833$), Glioblastoma multiforme dihydropyrimidinase-related protein 2 (DPYSl2) and/or methylenetetrahydrofolate dehydrogenase l (MTHFD1) and Glioblastoma multiforme collapsin response mediator protein l  (CRMP1) (pairwise distance $= 0.0.236014$). For these pairs, if a nanobody binds with one of them, we assume it also binds to the other. Thus, we add $1388$ of additional binding pairs that bind with each other. 

The non-binding pairs are obtained as follows: Suppose we have two binding Nb-Ag $(n_i, g_j)$ and $(n_{\ell}, g_k)$. Then both the pairs $(n_i, g_k)$ and $(n_{\ell}, g_j)$ are candidates for being declared as non-binding pairs if the distance between $g_i$ and $g_j$ is more than a certain threshold (we set the threshold $\in \{.8, .85, .9\}$. A random sample of such candidate pairs is added to the dataset as non-binding pairs, which consist of $1728$ such pairs in total. 

\paragraph{\textbf{Data Visualization:}}



\begin{figure}[h!]
  \centering
  \begin{subfigure}{.5\textwidth}
  \centering
  \includegraphics[scale=0.112]{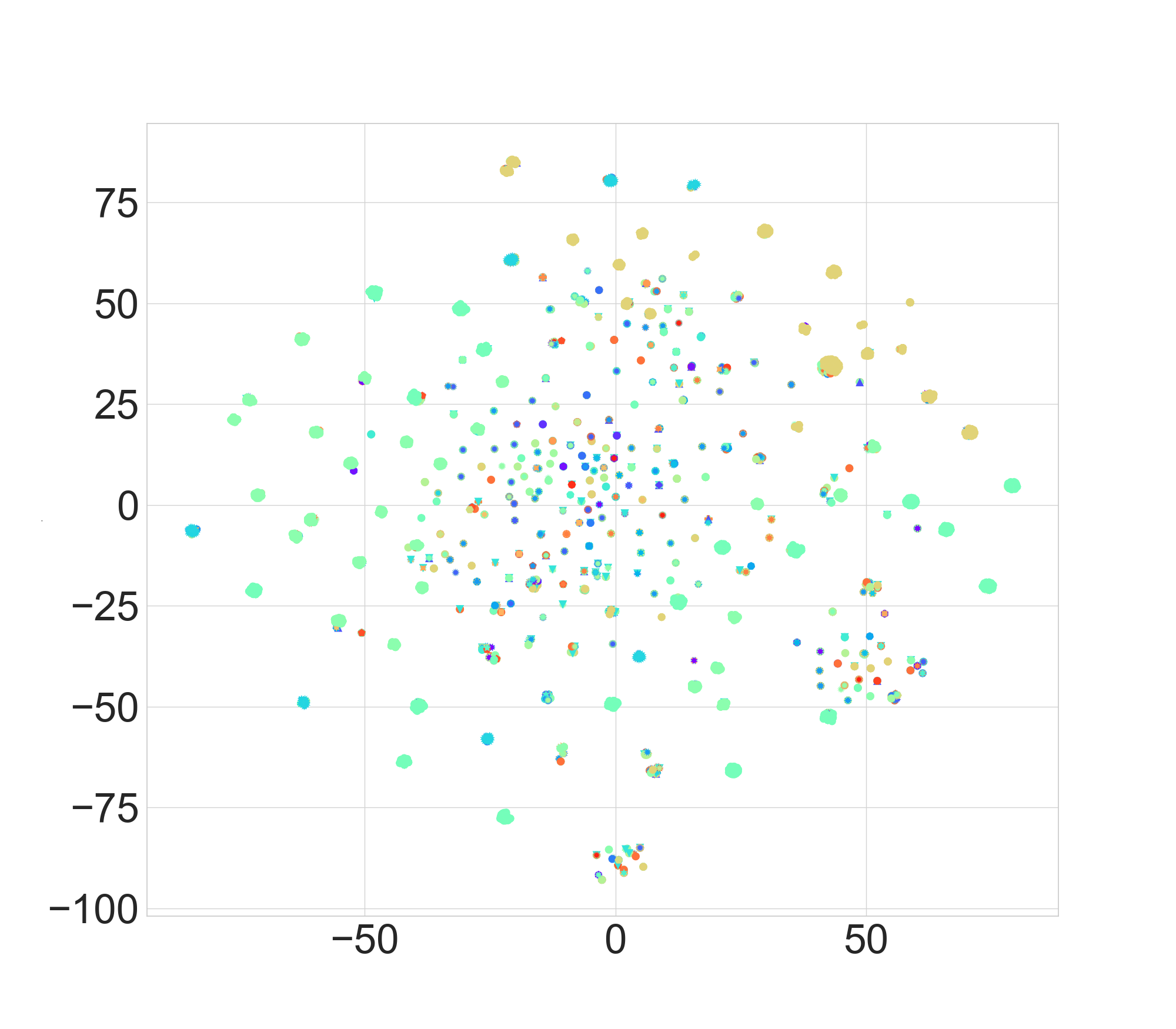}
  \caption{Nanobody}
  \end{subfigure}%
  \begin{subfigure}{.5\textwidth}
  \includegraphics[scale=0.112]{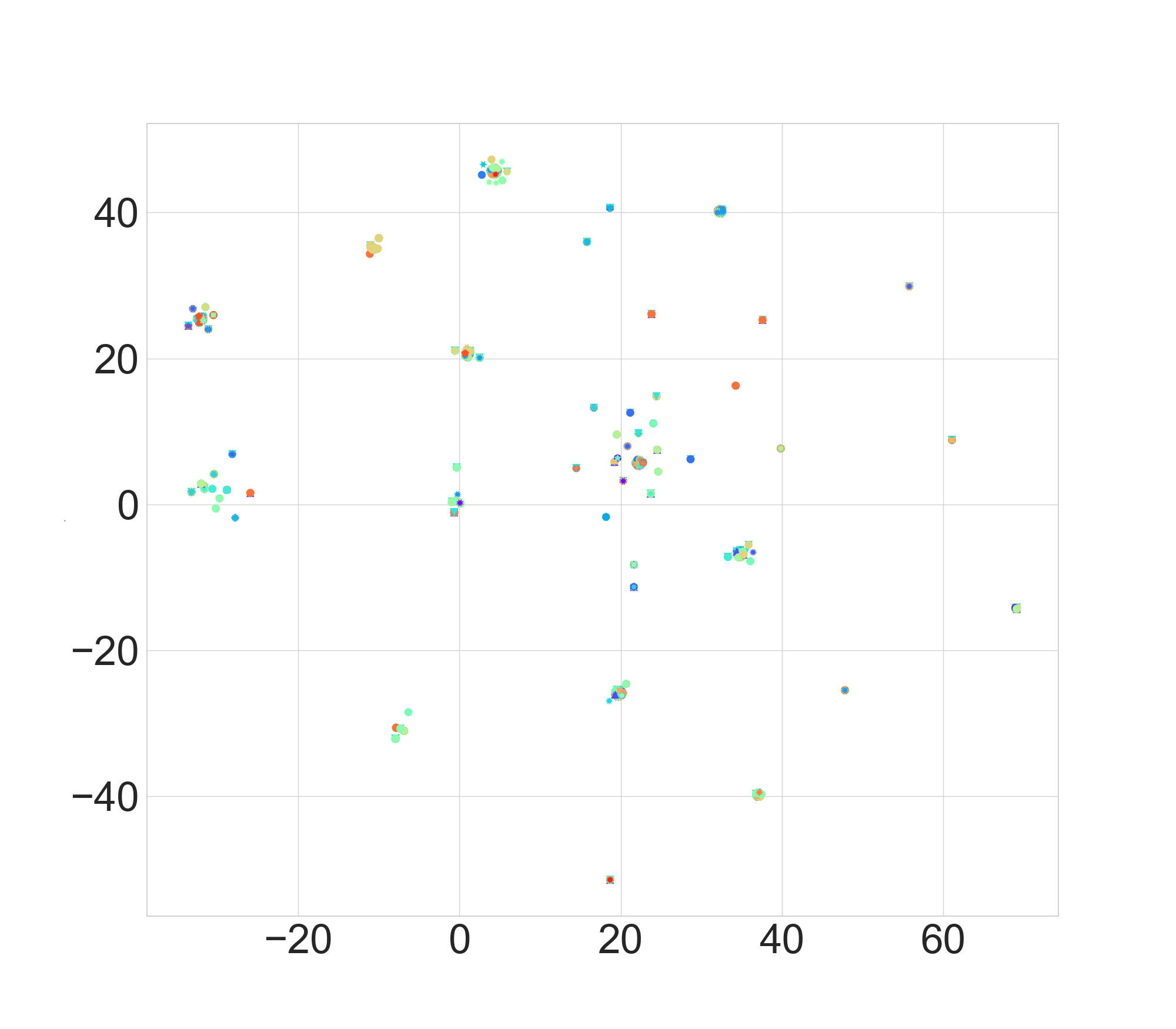}
  \caption{Antigen}
  \end{subfigure}%
  \caption{t-SNE plots for Nanobody and Antigen Embeddings using Spike2Vec approach. The figure is best seen in color.} 
  \label{fig_tsne}
\end{figure}

In order to visually assess the proximity of similar points in the Spike2Vec-based embeddings, we employ the t-Distributed Stochastic Neighbor Embedding (t-SNE) technique to obtain two-dimensional representations of the embeddings. These representations are then plotted using a scatterplot~\cite{van2008visualizing}. The t-SNE plots for the Nanobody and Antigen-based embeddings are depicted in Figure~\ref{fig_tsne}. The colored data points show the different antigen categories ($47$ in total).
Although the data points are scattered in the whole plot, we can observe small grouping for different labels, which shows that the embedding captures the hidden hierarchical and structural information inherent in the protein sequences.

\section{Representation Learning For Nb-Ag Binding Prediction}\label{sec_embedding_generation}
We learn various machine learning models on the training data of binding and non-binding pairs. To train the classifiers for the binary classification problem, we generate the feature vector (also called embeddings) for the nanobody and antibody in pairs. The corresponding feature vectors are a concatenation of the features extracted from the sequence as outlined above and feature vector embedding of the whole nanobody and antigen sequences, using state-of-the-art sequence2vec models discussed below. 

\subsubsection{\textbf{Embedding Generation:}}
To generate the fixed length numerical embedding from the variable length protein sequences, we use the idea of gapped $k$-mers. The gapped $k$-mers are a representation of biological sequences, such as DNA or protein sequences, that capture patterns while allowing for flexibility in the positioning of the $k$-mer elements (i.e., nucleotides or amino acids). In gapped $k$-mers, there are gaps or missing positions between the elements of the $k$-mers. These gaps introduce variability and enable the capture of more diverse and complex patterns compared to traditional contiguous $k$-mers.

The spectrum generated from gapped $k$-mers refers to the collection of all possible (unique) gapped $k$-mers (along with their count) that can be formed from a given protein sequence. These $k$-mers counts are then added into the numerical vector, which is generated based on all possible $k$-mers for the given alphabet $\Sigma$ that corresponds to \textit{ACDEFGHIKLMNPQRSTVWXY-}. Note that in $\Sigma$, the character '-' is used to include the gap in the $k$-mers. For example, for a $k$-mer 'ACD', the gapped $k$-mers will comprise of '-CD', 'A-D', 'AC-', and 'ACD'. An important point to note here is that we generated $4$ $k$-mers from just $1$ original $k$-mer. This extra information helps us to preserve more information, which eventually helps in the downstream supervised analysis. The pseudocode for generating gapped $k$-mers-based spectrum is given in Algorithm~\ref{algo_gapped_kmers}.
The spectrum provides a comprehensive representation of the sequence, taking into account both conserved elements and the flexibility introduced by the gaps. Each gapped $k$-mer in the spectrum represents a distinct pattern or subsequence present in the sequence, capturing variations and relationships between elements that may be important for various biological processes. Using gapped $k$-mers offers several advantages compared to other typical biological methods:

\begin{algorithm}[h!]
  \caption{Gapped $k$-mers Spectrum}
    \label{algo_gapped_kmers}
	\begin{algorithmic}[1]
 \scriptsize
	\State \textbf{Input:} Set of sequences $\mathcal{S}$, alphabet $\Sigma$, $k$ is size of $k$-mer
    \State \textbf{Output:} Spectrum $V$

    \State combos = GenerateAllCombinations($\Sigma$, k)  \Comment{all possible combinations of $k$-mers}
    \State $V$ = [0] * $\vert \Sigma \vert^{k}$  
    \For{ s in $\mathcal{S}$}
        \State $Gkmers$ = \Call{GappedKmers}{s,k} \Comment{compute all possible gapped $k$-mers}
        \For{ $i$ $\leftarrow 1$ to $ \vert Gkmers \vert$}
          \State idx = combos.index($Gkmers$[i])           
          \State $V$[idx] $\leftarrow$ $V$[idx] + 1             
    \EndFor
    \EndFor
    \State return($V$)
  \end{algorithmic}
\end{algorithm}

\paragraph{\textbf{Increased sensitivity: }} Gapped $k$-mers can capture more complex patterns and relationships compared to traditional contiguous $k$-mers. Gapped $k$-mers can capture conserved elements that are not necessarily adjacent in the sequence by allowing gaps. This increased sensitivity can be crucial for identifying motifs or regions of interest but may exhibit variable spacing.

\paragraph{\textbf{Enhanced flexibility: }} Gapped $k$-mers offer flexibility in terms of the spacing between elements. This flexibility allows for the inclusion of different variations and insertions, providing a more comprehensive representation of the sequence. Gapped $k$-mers can accommodate diverse patterns and handle insertions or deletions more effectively than traditional contiguous $k$-mers.

\paragraph{\textbf{Comprehensive motif representation: }} The spectrum generated from gapped $k$-mers provides a comprehensive representation of the sequence by capturing a wide range of conserved patterns and variations. This allows for a more detailed analysis of complex motifs or functional regions that involve specific arrangements of elements.

\paragraph{\textbf{Improved specificity: }} Gapped $k$-mers can help improve specificity by reducing false-positive matches. By considering both conserved elements and gaps, gapped $k$-mers can differentiate between true motifs and random matches that may occur by chance in traditional $k$-mers.


After generating the embeddings using gapped $k$-mers spectrum, we use those embeddings as input to the machine learning classifiers for binary classification to predict the binding of Nb-Ag pairs binding.

\section{Experimental Setup} \label{experimental_setup}
In this section, we discuss the dataset statistics and evaluation metrics along with baseline models.
All experiments are carried out using Python on a system equipped with a 2.4 GHz Core i5 processor, $32$ GB of memory, and the Windows 10 operating system. For experiments, we randomly divide the data into $70-30\%$ training and testing set and use $10\%$ of the data from the training set as the validation data. The experiments are repeated $5$ times and we show average results to eliminate any biases in the random splits.
For the sake of reproducibility, our code and preprocessed datasets can be accessed online~\footnote{\url{https://github.com/sarwanpasha/Nanobody_Antigen}}.

\paragraph{\textbf{Baseline Models:}}
To evaluate the proposed embedding, we compare their results with several popular baseline models from the literature. The baselines includes Spike2Vec~\cite{ali2021spike2vec}, Minimizers~\cite{roberts2004reducing}, and PWM2Vec~\cite{ali2022pwm2vec}. A detailed description of the baseline embedding models is given in Section~\ref{ap_sec_baseline} in the appendix.

\paragraph{\textbf{Evaluation Metrics:}}
To assess the performance of embeddings, we employ several evaluation metrics, including accuracy, precision, recall, F1 score (weighted), F1 score (macro), Receiver Operating Characteristic (ROC) curve, Area Under the Curve (AUC), and training runtime. For metrics designed for binary classification tasks, we adopt the one-vs-rest approach for multi-class classification.

\paragraph{\textbf{Machine Learning Classifiers:}}
Supervised analysis entails the utilization of diverse linear and non-linear classifiers, such as Support Vector Machine (SVM), Naive Bayes (NB), Multi-Layer Perceptron (MLP), K-Nearest Neighbors (KNN), Random Forest (RF), Logistic Regression (LR), and Decision Tree (DT).

\section{Results And Discussion} \label{results_discussion}
The classification results for different evaluation metrics with and without the sequence features are reported in Table~\ref{tbl_results_classification}. We observe (bold values) that the gapped $k$-mers spectrum outperforms all embeddings in the case of average accuracy, precision, recall, and ROC-AUC using the random forest classifier. For Weighted and Macro F1, the baseline Spike2Vec performs better than other embeddings using the random forest classifier. One interesting observation is that the random forest classifier consistently outperforms other classifiers for all embeddings and evaluation metrics, as shown with underlined values in Table~\ref{tbl_results_classification}.

\begin{table}[h!]
\centering
\resizebox{0.86\textwidth}{!}{
 \begin{tabular}{@{\extracolsep{6pt}}p{1.5cm}p{1.5cm}lp{1.1cm}p{1.1cm}p{1.1cm}p{1.3cm}p{1.5cm}p{1.1cm}p{1.7cm}}
    \toprule
        & \multirow{2}{*}{Embeddings} & \multirow{2}{*}{Algo.} & \multirow{2}{*}{Acc. $\uparrow$} & \multirow{2}{*}{Prec. $\uparrow$} & \multirow{2}{*}{Recall $\uparrow$} & \multirow{2}{1.7cm}{F1 (Weig.) $\uparrow$} & \multirow{2}{1.7cm}{F1 (Macro) $\uparrow$} & \multirow{2}{1.2cm}{ROC AUC $\uparrow$} & Train Time (sec.) $\downarrow$\\
        \midrule \midrule
        \multirow{28}{1.4cm}{Without Sequence Features} & \multirow{7}{1.2cm}{Spike2Vec}
         & SVM & 0.818 & 0.824 & 0.818 & 0.818 & 0.818 & 0.819 & 5.662 \\
         & & NB & 0.813 & 0.815 & 0.813 & 0.813 & 0.813 & 0.813 & \underline{0.103} \\
        &  & MLP & 0.844 & 0.846 & 0.844 & 0.844 & 0.844 & 0.844 & 4.075 \\
         & & KNN & 0.892 & 0.893 & 0.892 & 0.892 & 0.892 & 0.892 & 1.290 \\
         & & RF & \underline{0.906} & \underline{0.911} & \underline{0.906} & \underline{\textbf{0.906}} & \underline{\textbf{0.906}} & \underline{0.906} & 3.725 \\
         & & LR & 0.813 & 0.815 & 0.813 & 0.813 & 0.813 & 0.814 & 2.417 \\
         & & DT & 0.878 & 0.878 & 0.878 & 0.878 & 0.877 & 0.878 & 1.293 \\
         \cmidrule{2-10} 
          & \multirow{7}{1.2cm}{Minimizers} & SVM & 0.824 & 0.826 & 0.824 & 0.823 & 0.823 & 0.823 & 5.444 \\
         &  & NB & 0.791 & 0.792 & 0.791 & 0.790 & 0.790 & 0.790 & \underline{0.091} \\
         &  & MLP & 0.844 & 0.845 & 0.844 & 0.844 & 0.844 & 0.844 & 2.997 \\
         &  & KNN & 0.880 & 0.880 & 0.880 & 0.880 & 0.880 & 0.880 & 1.257 \\
         &  & RF & \underline{0.892} & \underline{0.898} & \underline{0.892} & \underline{0.892} & \underline{0.892} & \underline{0.893} & 4.000 \\
         &  & LR & 0.811 & 0.812 & 0.811 & 0.811 & 0.811 & 0.811 & 1.343 \\
         &  & DT & 0.851 & 0.851 & 0.851 & 0.850 & 0.850 & 0.850 & 1.677 \\
          \cmidrule{2-10} 
         & \multirow{7}{1.2cm}{PWM2Vec} 
          & SVM & 0.810 & 0.812 & 0.810 & 0.809 & 0.809 & 0.809 & 5.732 \\
         &  & NB & 0.792 & 0.793 & 0.792 & 0.792 & 0.792 & 0.792 & \underline{0.095} \\
         &  & MLP & 0.820 & 0.821 & 0.820 & 0.820 & 0.819 & 0.820 & 3.730 \\
         &  & KNN & 0.875 & 0.875 & 0.875 & 0.875 & 0.875 & 0.875 & 1.232 \\
         &  & RF & \underline{0.892} & \underline{0.899} & \underline{0.892} & \underline{0.891} & \underline{0.891} & \underline{0.892} & 3.746 \\
         &  & LR & 0.804 & 0.805 & 0.804 & 0.804 & 0.804 & 0.804 & 7.137 \\
         &  & DT & 0.866 & 0.866 & 0.866 & 0.866 & 0.866 & 0.866 & 1.692 \\
          \cmidrule{2-10} 
         & \multirow{7}{1.2cm}{Gapped $k$-mers} 
          & SVM & 0.814 & 0.816 & 0.814 & 0.813 & 0.813 & 0.812 & 5.740 \\
         &  & NB & 0.798 & 0.798 & 0.798 & 0.797 & 0.797 & 0.796 & \underline{0.087} \\
         &  & MLP & 0.824 & 0.825 & 0.824 & 0.824 & 0.824 & 0.824 & 2.886 \\
         &  & KNN & 0.885 & 0.886 & 0.885 & 0.885 & 0.885 & 0.885 & 0.995 \\
         &  & RF & \underline{\textbf{0.907}} & \underline{\textbf{0.912}} & \underline{\textbf{0.907}} & \underline{0.894} & \underline{0.894} & \underline{\textbf{0.908}} & 3.755 \\
         &  & LR & 0.812 & 0.813 & 0.812 & 0.812 & 0.812 & 0.812 & 4.395 \\
         &  & DT & 0.872 & 0.872 & 0.872 & 0.872 & 0.871 & 0.872 & 1.777 \\
        \midrule
         \multirow{28}{1.4cm}{With Sequence Features} & \multirow{7}{1.2cm}{Spike2Vec}
          & SVM & 0.791 & 0.796 & 0.791 & 0.790 & 0.790 & 0.790 & 8.804 \\
         &  & NB & 0.695 & 0.737 & 0.695 & 0.680 & 0.678 & 0.691 & \underline{0.085} \\
         &  & MLP & 0.811 & 0.814 & 0.811 & 0.811 & 0.811 & 0.811 & 2.326 \\
         &  & KNN & 0.844 & 0.845 & 0.844 & 0.844 & 0.844 & 0.844 & 0.953 \\
         &  & RF & \underline{0.897} & \underline{0.903} & \underline{0.897} & \underline{0.896} & \underline{0.896} & \underline{0.898} & 3.890 \\
         &  & LR & 0.827 & 0.827 & 0.827 & 0.827 & 0.826 & 0.827 & 1.183 \\
         &  & DT & 0.847 & 0.848 & 0.847 & 0.847 & 0.847 & 0.847 & 1.246 \\
         \cmidrule{2-10} 
          & \multirow{7}{1.2cm}{Minimizers} & SVM & 0.778 & 0.783 & 0.778 & 0.777 & 0.777 & 0.777 & 10.938 \\
        &  & NB & 0.674 & 0.736 & 0.674 & 0.649 & 0.647 & 0.670 & \underline{0.094} \\
         &  & MLP & 0.801 & 0.806 & 0.801 & 0.800 & 0.800 & 0.800 & 3.228 \\
         &  & KNN & 0.842 & 0.842 & 0.842 & 0.842 & 0.842 & 0.842 & 0.827 \\
         &  & RF & \underline{0.896} & \underline{0.902} & \underline{0.896} & \underline{0.896} & \underline{0.896} & \underline{0.897} & 3.801 \\
         &  & LR & 0.823 & 0.823 & 0.823 & 0.823 & 0.823 & 0.823 & 1.167 \\
         &  & DT & 0.846 & 0.846 & 0.846 & 0.845 & 0.845 & 0.845 & 1.297 \\
  \cmidrule{2-10} 
         & \multirow{7}{1.2cm}{PWM2Vec} 
          & SVM & 0.766 & 0.770 & 0.766 & 0.765 & 0.765 & 0.766 & 9.569 \\
         &  & NB & 0.679 & 0.726 & 0.679 & 0.659 & 0.657 & 0.674 & \underline{0.087} \\
         &  & MLP & 0.811 & 0.813 & 0.811 & 0.811 & 0.811 & 0.811 & 2.889 \\
         &  & KNN & 0.828 & 0.828 & 0.828 & 0.827 & 0.827 & 0.827 & 0.768 \\
         &  & RF & \underline{0.893} & \underline{0.901} & \underline{0.893} & \underline{0.892} & \underline{0.892} & \underline{0.894} & 3.765 \\
         &  & LR & 0.819 & 0.819 & 0.819 & 0.819 & 0.819 & 0.819 & 1.495 \\
         &  & DT & 0.851 & 0.851 & 0.851 & 0.851 & 0.851 & 0.850 & 1.279 \\
 \cmidrule{2-10} 
         & \multirow{7}{1.2cm}{Gapped $k$-mers} 
        & SVM & 0.785 & 0.792 & 0.785 & 0.784 & 0.783 & 0.784 & 9.270 \\
         &  & NB & 0.720 & 0.745 & 0.720 & 0.712 & 0.711 & 0.718 & \underline{0.086} \\
         &  & MLP & 0.807 & 0.810 & 0.807 & 0.806 & 0.806 & 0.806 & 2.432 \\
         &  & KNN & 0.839 & 0.839 & 0.839 & 0.839 & 0.838 & 0.838 & 0.753 \\
         &  & RF & \underline{0.895} & \underline{0.901} & \underline{0.895} & \underline{0.894} & \underline{0.894} & \underline{0.895} & 3.468 \\
         &  & LR & 0.823 & 0.823 & 0.823 & 0.823 & 0.823 & 0.823 & 1.123 \\
         &  & DT & 0.860 & 0.861 & 0.860 & 0.860 & 0.860 & 0.860 & 0.955 \\

         \bottomrule
         \end{tabular}
}
 \caption{Classification results (averaged over $5$ runs) for different evaluation metrics. The best values for each embedding are underlined while overall best values among all embeddings for different evaluation metrics are shown in bold.}
    \label{tbl_results_classification}
\end{table}

Comparing embeddings with and without sequence features, we observe that using these features degrades classifiers' accuracy. This degradation is due to redundancy and feature dimensionality as some of the features might capture similar information as the $k$-mers spectrum. For example, the charge at pH, aromaticity, or GRAVY may already encode certain aspects of protein sequence patterns. Including redundant features can lead to multicollinearity, where the features are highly correlated, making it difficult for the classifier to distinguish their individual contributions. In the case of feature dimensionality, the addition of new features increases the dimensionality of the input space. With a higher number of features, the classifier faces the curse of dimensionality. Insufficient training data or a limited number of samples in each class relative to the feature space can result in overfitting, reduced generalization performance, and decreased accuracy.

\subsection{Statistical Significance}
Considering the random splitting of the data into training and testing sets during the computation of the results, the statistical significance of the obtained results becomes a significant concern. To address this, we conducted a student t-test to evaluate the significance of the results. The $p$-values were determined using the averages (as shown in the results table) and standard deviations (SD) derived from five different runs.
It is worth noting that the SD values obtained from these five runs, which involved different random splits, were consistently small. In the majority of cases, they were less than 0.002. Consequently, we observed that the $p$-values were predominantly less than 0.05 due to the low SD values. This confirms the statistical significance of the obtained results.
Due to page limitation, we did not include the SD values and $p$-values. However, it is important to highlight that the small SD values and the majority of $p$-values being less than 0.05 further support the statistical significance of our results.

\section{Conclusion}\label{sec_conclusion}
This study aimed to develop an ML approach to predict Nb-Ag binding solely based on sequences, thereby reducing the need for computationally intensive techniques such as docking. The proposed method utilized an embedding approach using gapped $k$-mers to generate a spectrum, which was then used for supervised analysis. Experimental evaluation of our approach demonstrates that the gapped $k$-mers spectrum outperformed competing embeddings. Our approach offers a more efficient and cost-effective alternative for screening potential Nbs and holds promise for facilitating the development of Nb-based diagnostics and therapeutics for various diseases, including cancer and other serious illnesses. Future research involves evaluating the proposed model on bigger datasets and also working on the generalizability and robustness of the model. Additionally, exploring the integration of additional features and considering other machine learning algorithms could further enhance predictive performance.

\bibliographystyle{splncs04}
\bibliography{references}

\clearpage


\appendix

\section*{Appendix}

\section{Baseline Models}\label{ap_sec_baseline}

\paragraph{\textbf{Spike2Vec~\cite{ali2021spike2vec}: }}
The objective of this approach is to generate numerical embeddings for input protein sequences, facilitating the utilization of machine learning models. To achieve this, the method begins by creating $k$-mers from the given spike sequence. $k$-mers are employed due to their ability to retain the sequential information of the protein sequence.

\begin{definition}[$k$-mers]\label{def_kmers}
$K$-mers refer to sets of consecutive amino acids (called mers) of length $k$ in a given sequence. The consecutive $k$-mers are computed using the sliding window approach, where the next $k$-mer is 1 window to the right of the previous $k$-mer. In the domain of natural language processing (NLP), they are referred to as n-grams.
\end{definition}
All possible $k$-mers that can be generated for a given sequence of length $N$ are $N - k + 1$.
Spike2Vec calculates the frequency vector based on $k$-mers to convert the alphabetical information of $k$-mers into a numerical representation. This vector captures the occurrence counts of each $k$-mer in the sequence, also called $k$-mers spectrum.

\paragraph{\textbf{Minimizers~\cite{roberts2004reducing}: }}
The Minimizer-based feature vector is a method that involves computing a \emph{minimizer} of length $m$ (also called $m$-mer) for a given $k$-mer. In the case of $m$-mer, we have $m < k$. The $m$-mer is determined as the lexicographically smallest sequence in both the forward and reverse order of the $k$-mer.
A fixed-length frequency vector is constructed from the set of minimizers, where each bin in the frequency vector represents the count of a specific minimizer. This method is also referred to as $m$-mers spectrum. The length of each vector is determined by the alphabet size $\Sigma$ (where $\Sigma$ contains all possible characters or amino acids in protein sequence i.e. \textit{ACDEFGHIKLMNPQRSTVWXY}) and the length of the minimizers denoted as $m$ (we set $m$ to 3, which is decided using the standard validation set approach). Hence, the length of the vector is $\vert \Sigma \vert^m$.

\paragraph{\textbf{PWM2Vec~\cite{ali2022pwm2vec}: }}
The PWM2Vec is a feature embedding method that transforms protein sequences into numerical representations. Instead of relying on the frequency of $k$-mers, PWM2Vec assigns weights to each amino acid within a $k$-mer. These weights are determined using a position weight matrix (PWM) associated with the $k$-mer. By considering the position-specific importance of amino acids, PWM2Vec captures both the relative significance of amino acids and preserves the ordering information within the sequences.

\end{document}